\documentstyle[mprocl,
psfig]{article}

\def\Journal#1#2#3#4{{#1} {\bf #2}, #3 (#4)}

\def\JHEP{\em JHEP}
\def\PHY{\em Physica}
\def\AP{\em Ann. Phys.}

\def\NPB{{\em Nucl. Phys.} B}
\def\PLA{{\em Phys. Lett.}  A}
\def\PLB{{\em Phys. Lett.}  B}
\def\PRL{\em Phys. Rev. Lett.}
\def\PRD{{\em Phys. Rev.} D}
\def\PRB{{\em Phys. Rev.} B}

\input{psfig.sty}
\def\Journal#1#2#3#4{{#1} {\bf #2}, #3 (#4)}

\def\NPB{{\em Nucl. Phys.} B}
\def\PLB{{\em Phys. Lett.}  B}
\def\PRL{\em Phys. Rev. Lett.}
\def\PRD{{\em Phys. Rev.} D}

\def\be{\begin{equation}}
\def\ee{\end{equation}}
\def\bea{\begin{eqnarray}}
\def\eea{\end{eqnarray}}

\def \lta {\mathrel{\vcenter
     {\hbox{$<$}\nointerlineskip\hbox{$\sim$}}}}
\def \gta {\mathrel{\vcenter
     {\hbox{$>$}\nointerlineskip\hbox{$\sim$}}}}

\newcommand{\rt}{\tilde{r}}
\newcommand{\Ut}{\tilde{U}}

\newcommand{\At}{\tilde{A}}
\newcommand{\St}{\tilde{S}}

\newcommand{\phibounce}{\phi_b}

\newcommand{\phit}{\tilde{\phi}}

\newcommand{\dxl}{\delta x_{\Lambda}}
\newcommand{\dkl}{\delta \kappa_{\Lambda}}
\newcommand{\dkx}{\delta x_{\Lambda}}

\newcommand{\lx}{\lambda}
\newcommand{\Lx}{\Lambda}
\newcommand{\ex}{\epsilon}

\newcommand{\mt}{\tilde{m}}

\newcommand{\rhoa}{\rho_1}
\newcommand{\rhob}{\rho_2}

\newcommand{\rht}{\tilde{\rho}}
\newcommand{\rhta}{\tilde{\rho}_1}
\newcommand{\rhtb}{\tilde{\rho}_2}

\newcommand{\kx}{\kappa}

\newcommand{\een}{\end{subequations}}
\newcommand{\ben}{\begin{subequations}}
\newcommand{\beq}{\begin{eqalignno}}
\newcommand{\eeq}{\end{eqalignno}}

\begin{document}

\title{THE EXACT RENORMALIZATION GROUP 
AND 
FIRST-ORDER PHASE TRANSITIONS}

\author{N. TETRADIS}

\address{Department of Physics, University of Crete, 710 03 Heraklion,
Crete, Greece \\and\\
Department of Physics, University of Athens, 157 71 Athens, Greece}

\maketitle\abstracts{ 
Studies of first-order phase transitions through the
use of the exact renormalization group are reviewed. 
In the first part the emphasis is on universal aspects: 
We discuss the universal critical behaviour near
weakly first-order phase transitions for a three-dimensional 
model of two coupled scalar fields -- the cubic anisotropy model.
In the second part we review the application of the exact
renormalization group to the calculation of bubble-nucleation rates.
More specifically, we concentrate on the pre-exponential factor.
We discuss the reliability of homogeneous nucleation theory
that employs a saddle-point expansion around the critical
bubble for the calculation of the nucleation rate. 
}

\section{Weakly first-order phase transitions}

Weakly first-order phase transitions appear in 
many statistical systems
(such as superconductors or anisotropic systems)
and in high-temperature field theories. 
Many cosmological phase transitions, such as the electroweak \cite{review},  
fall in this category. However, a reliable quantitative description
of these phenomena is still lacking.

A simple example of a system with an arbitrarily weakly 
first-order phase transition is the cubic anisotropy model
\cite{rudnick,aharony}. In field-theoretical language,  
it corresponds to 
a theory of two real scalar fields 
$\phi_a~(a=1,2)$ in three dimensions,  
invariant under the discrete symmetry
$(1 \leftrightarrow - 1,
2 \leftrightarrow - 2,
1 \leftrightarrow  2)$. 
It can also be considered as the effective description of
the four-dimensional high-temperature theory with the same
symmetry, at energy scales smaller than the temperature
\cite{stefan}. 

The properties of the weakly 
first-order phase transitions in this model have been studied 
within the $\epsilon$-expansion \cite{arnolde} or through
Monte Carlo simulations \cite{arnoldl}. In particular,
universal amplitudes have been computed, which describe 
the relative discontinuity
of various physical quantities along the phase transition, in the 
limit when the transition becomes arbitrarily weakly first order.
A discrepancy has been observed in the predictions for the 
universal ratio of susceptibilities $\chi_+/\chi_-$ on either
side of the phase transition, obtained through
Monte Carlo simulations or the $\epsilon$-expansion
\cite{arnolds}. The Monte Carlo simulations
predict $\chi_+/\chi_-=4.1(5)$, while the first three orders of the
$\ex$-expansion give $\chi_+/\chi_-=2.0, 2.9, 2.3$ respectively.

We summarize here the results of an alternative approach \cite{weak} 
that employs the 
exact renormalization group.
It is based on the effective average action $\Gamma_k$ 
\cite{average}, which is a coarse-grained
free energy with an infrared cutoff. More precisely, 
$\Gamma_k$  incorporates the effects
of all fluctuations with momenta $q^2 > k^2$, but not those with
$q^2 < k^2$. In the limit
$k \rightarrow 0$, the effective average action 
becomes the standard effective action
(the generating functional of the 1PI Green functions), while at a high
momentum scale (of the order of the ultraviolet cutoff) 
$k =\Lx \rightarrow \infty$,
it equals the classical or bare
action.
An exact non-perturbative flow equation 
determines the scale dependence of
$\Gamma_k$.

The flow equation is a functional differential
equation, and an approximate solution requires a truncation. 
Our truncation is the lowest order in a systematic derivative
expansion of $\Gamma_k$ \cite{indices,morris}
\be
\Gamma_k = \int d^3 x \left\{
U_k(\rho_1,\rho_2) + \frac{1}{2} Z_k 
\left( \partial^{\mu} \phi_1 \partial_{\mu} \phi^1 
+ \partial^{\mu} \phi_2 \partial_{\mu} \phi^2 
\right) \right\}.
\label{two} \ee
Here  
$\rho_a = \frac{1}{2} \phi_a \phi^a$ and 
the potential $U_k(\rho_1,\rho_2)$ is symmetric under the interchange
$1 \leftrightarrow 2$.
The wave-function renormalization
is approximated by one $k$-dependent parameter $Z_k$. 
The truncation of the higher derivative terms in the action  
is expected to generate an uncertainty of the order of the anomalous
dimension $\eta$. 
For the model we are considering, 
$\eta \simeq 0.035$ and the induced error is small. 

The fixed-point structure of the theory 
is more easily identified if
we use the dimensionless renormalized parameters
\be
\rht_a= ~Z_k k^{-1} \rho_a
~~~~~~~~~~~~~~
u_k(\rht_1,\rht_2) = ~k^{-3} U_k(\rhoa,\rhob). 
\label{three}
\ee
The evolution equation for the potential can now be written in
the scale-independent form \cite{stefan}
\be
\frac{\partial}{\partial t} u_k(\rhta,\rhtb) =
-3 u_k +(1 + \eta) (\rht_1 u_1 + \rht_2 u_2) 
- \frac{1}{8 \pi^2} L^3_0(\mt^2_1)
- \frac{1}{8 \pi^2} L^3_0(\mt^2_2),
\label{four} \ee
where $t=\ln(k/\Lx)$.
The anomalous dimension $\eta$ is defined as 
$d \ln Z_k/dt = -\eta$. For the part of the phase diagram of interest,
it is constant, $\eta \simeq 0.035$, to a good approximation \cite{stefan}. 
The quantities
$\mt^2_{1,2}$ are the eigenvalues of the rescaled mass matrix at the
point $(\rhta,\rhtb)$ 
\be
2 \mt^2_{1,2} = 
 u_1 + u_2 +2 u_{11} \rhta + 2 u_{22} \rhtb 
\pm  \sqrt{(u_1 - u_2 +2 u_{11} \rhta - 2 u_{22} \rhtb )^2 
+ 16 u^2_{12} \rhta \rhtb }.
\label{five} \ee
We use the
notation $u_1 = {\partial u_k}/{\partial \rhta}$,
$u_{12} = {\partial^2 u_k}/{\partial \rhta \partial \rhtb}$, etc. 
The function $L^3_0(w)$
is negative for all values of $w$. 
Also $|L^3_n(w)|$ is monotonically decreasing for increasing $w$
and introduces a threshold behaviour in the evolution. 
For large values of $\mt^2_a$
the last two terms in eq. (\ref{four}) vanish and the evolution
of $U_k$ stops \cite{average,indices}.

The initial condition for the integration is provided by the
bare potential, which is identified with the
effective average potential at a very high scale $k=\Lx$. 
We use a bare potential of the form 
\be
 u_{\Lx}(\rhta,\rhtb) = 
\frac{1}{2} {\lx}_{\Lx} \left\{ \left(\rhta-\kx_{\Lx}\right)^2 
+  \left(\rhtb-\kx_{\Lx}\right)^2 \right\} 
+ (1+x_{\Lx})  {\lx}_{\Lx} \rhta \rhtb
\label{twob} \ee
and $Z_{\Lx}=1$.

The phase diagram of the theory \cite{rudnick,aharony,stefan}
has three fixed points that govern the dynamics 
of second-order phase transitions. 
They are located on the critical surface separating the 
phase with symmetry breaking from the symmetric one.
The most stable of them corresponds to a system with 
an increased $O(2)$ symmetry. It can be approached directly from a bare action 
given by eq. (\ref{twob}) with $x_{\Lx}=0$.
The other two 
are Wilson--Fisher fixed points, corresponding to two disconnected 
$Z_2$-symmetric theories. One of them can be approached from a bare action with
$x_{\Lx}=-1$, while the second requires $x_{\Lx}=2$ 
\footnote{
A redefinition of the
fields demonstrates that this choice corresponds to two 
disconnected $Z_2$-symmetric theories \cite{stefan}.}.
Flows that start with $-1 < x_{\Lx} < 2$ and near the critical surface 
eventually lead to the $O(2)$-symmetric fixed point.
For $x_{\Lx} > 2$ or $x_{\Lx} < -1$ the evolution leads
to a region of first-order phase transitions. If $x_{\Lx}$ is chosen
slightly larger than 2 or slightly smaller than $-1$ the 
phase transitions are weakly first order. 
The evolution first approaches
one of the fixed points 
before a second minimum appears in the potential. 

\begin{figure}
\begin{center}
\hspace{0.cm}
\psfig{figure=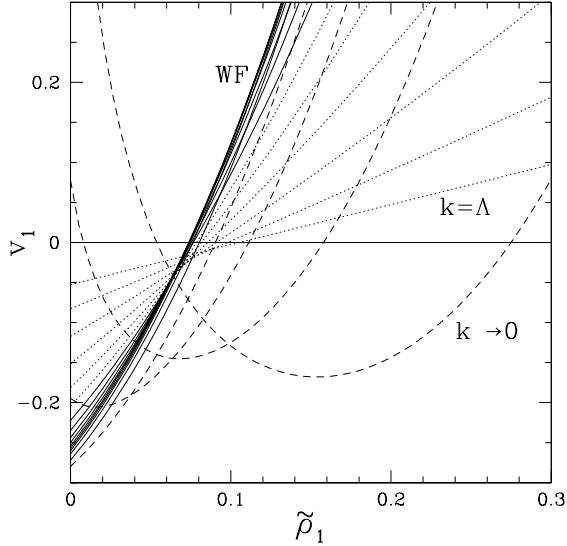,height=3.in}
\end{center}
\caption{
The derivative of the rescaled 
potential along the $\rho_1$ axis, as the coarse-graining scale
is lowered. 
\label{fig:one}}
\end{figure}

\begin{figure}
\begin{center}
\hspace{0.cm}
\psfig{figure=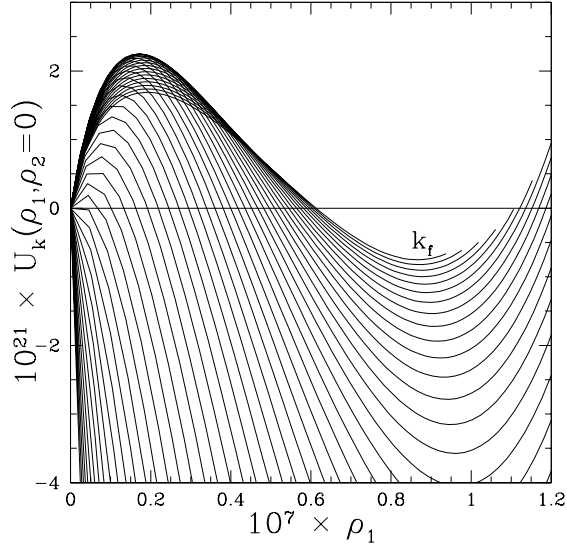,height=3.in}
\end{center}
\caption{The potential along the $\rho_1$ axis, as the 
coarse-graining scale is lowered.
\label{fig:two}}
\end{figure}

The partial differential 
equation (\ref{four}) with $\eta=0.035$ can be integrated 
numerically \cite{weak}.
We are interested in the region $x \ge 2$. 
The region $x \le -1$ can be mapped on it through a redefinition
of the fields \cite{stefan}.
We concentrate on the potential along the $\rhta$ axis and
we define $v_k(\rhta )=u_k(\rhta,\rhtb=0)$.
In fig. 1 we present 
the evolution of
$v_1 = \partial v_k / \partial \rhta$, 
starting at a very high scale $k=\Lx$
with a bare potential 
given by eq. (\ref{twob}).
All dimensionful quantities are normalized 
with respect to $\Lx$. The initial 
coupling $\lx_{\Lx}$ is chosen arbitrarily, while
the minimum $\kx_\Lx$ of the bare potential is
taken very close to the critical value $\kx_{cr}$ 
that separates the phase with symmetry breaking from the symmetric one.
For $|\dkl| = |\kx_\Lx - \kx_{cr}| \ll 1$ the system spends a long ``time''
$t$ of its evolution on the critical surface separating the two phases.
The initial value $x_{\Lx}$ 
is taken slightly larger than the fixed-point value 2.
After the initial evolution (dotted lines) the potential settles down
near the Wilson--Fisher fixed point (solid lines). This fixed point,
that has a mixing term between the two fields with
$x=2$, is repulsive in the $x$ direction. 
Eventually $x$ evolves towards larger
values. This forces the potential to move away from its scale-independent
form (dashed lines). At some point in the subsequent evolution the 
curvature of the potential at the origin $v_1(\rhta=0)$ becomes
positive. This signals the appearance of a new minimum there, and the 
presence of a radiatively-induced (or fluctuation-driven)
first-order phase transition. 
The evolution of the potential after it moves away from
the fixed point is
insensitive to the details of the bare potential. It is
uniquely determined by 
$|\dkl| = |\kx_\Lx - \kx_{cr}| \ll 1$ and $\dkx = x_\Lx-2 \ll 1$, 
and displays
{\bf universal} behaviour \cite{coarse}. 

In the limit $k \rightarrow 0$ and in the convex regions,
the rescaled potential $u_k$ 
grows and eventually diverges in such a way 
that $U_k$ becomes asymptotically
constant, equal to the effective potential $U=U_0$. This is apparent in
fig. 2, where the potential along the $\rho_1$ axis is plotted. 
The evolution of the non-convex part
of the potential (between the two minima)
is related to the issue of 
the convexity of the effective potential. This part should become flat for
$k \rightarrow 0$ \cite{coarse}. 
The approach to convexity is apparent in fig. 2, 
even though we have not followed 
the evolution all the way to $k=0$.

The relative magnitude of 
$|\dkl| = |\kx_\Lx - \kx_{cr}| \ll 1$ and $\dkx = x_\Lx-2 \ll 1$ 
results in different
types of evolution. For the type of behaviour depicted in figs. 1 and 2 
one must
take $|\dkl| \ll \dxl$. In the opposite limit, $|\dkl| \gg \dxl$,
the system leaves the critical surface before the mixing $x$ between the 
fields evolves 
away from the fixed-point value 2. As a result, a second minimum never
appears at the origin. Instead, the only minimum of the effective
potential $U_0$ is located
either at zero (symmetric phase) or away from it (phase with symmetry
breaking), depending on the sign of $\dkl$. 
The resulting phase transition is second
order and occurs for $\dkl=0$. 

We are interested in the universal ratio of susceptibilities 
$\chi_+/\chi_-$ on either
side of the phase transition. This ratio depends on the value
of $|\dkl| / \dkx$. For $|\dkl| / \dkx \gg 1$ we have 
$x  \simeq 2$ during
the whole evolution. The universal quantities characterizing the
second-order phase transition are determined by the Wilson--Fisher
fixed point. We calculate 
$\chi_+/\chi_-$ by integrating the evolution equation 
and evaluating 
$d^2 U_0/d \phi_1^2 = \chi^{-1}$ at the minimum, 
for $\dkl = \mp \epsilon$ with
$\epsilon \ll 1$.
We obtain $\chi_+/\chi_- = 4.1$. 
The evaluation of the same quantity through the $\epsilon$-expansion or
an expansion at fixed dimension gives 
$\chi_+/\chi_- = 4.8(3)$ \cite{zinn}, whereas 
experimental information gives $\chi_+/\chi_- = 4.3(3)$
\cite{zinn}.
The difference with the results of our method 
is due to the omission of higher derivative terms 
in the effective average action of eq. (\ref{two}) and approximations
that have been used for the numerical integration of eq. (\ref{four}).

For $|\dkl| / \dkx \ll 1$ the potential develops a second minimum at the
origin during
the later stages of the evolution. The phase transition is approached by 
fine tuning $\dkl$, so that the two minima have equal depth for $k=0$. 
The ratio of susceptibilities can be computed through the evaluation of
the second derivatives
of the potential at the two minima.  
Our final estimate is \cite{weak}
$\chi_+/\chi_-=1.7(7).$
It is in good agreement with the 
predictions of the
$\ex$-expansion 
\cite{arnolde,arnolds}
($\chi_+/\chi_-=2.0, 2.9, 2.3$ 
for the first three orders).
However, it disagrees with the lattice result 
\cite{arnoldl,arnolds} ($\chi_+/\chi_-=4.1(5)$).

In summary, the approach we presented
can provide a quantitative 
description of the universal behaviour near 
weakly first-order phase transitions.
It is based on the calculation of a coarse-grained free
energy through an exact flow equation. Fixed points in the evolution, 
the appearance of new minima in the potential, and the universal 
properties of the
resulting radiatively-induced first-order phase transitions 
can be studied in detail.

\renewcommand{\varphi}{\phi}

\section{Spontaneous nucleation and coarse graining}

\subsection{The necessity of coarse graining}

Near a first-order phase transition the potential (or free energy density)
of the system
has two separate local minima. As the temperature drops below a
critical value,
the minimum corresponding to the true vacuum
(for late times) becomes deeper than the one corresponding to
the false vacuum. However, the system may not adapt 
immediately to the new equilibrium situation, and we encounter
the familiar phenomena of supercooling or hysteresis. For sufficiently
low temperature the phase
transition takes place, driven by spontaneous nucleation. 
For example, as vapor
is cooled below the critical temperature, local droplets spontaneously form
and grow until the transition is completed. The inverse evolution
proceeds by the formation of vapor bubbles in a liquid. The
transition in ferromagnets is characterized by the formation
of local Weiss domains.
The formation of bubbles of the new vacuum is similar to
a tunnelling process, and typically exponentially suppressed
at the early stages of the transition. The reason is the
barrier between the local minima. The transition requires
the formation of 
the configuration with lowest action on the barrier. The rate includes
a Boltzmann factor involving the action of this
critical configuration.

Our present understanding of the phenomenon of nucleation
is based largely on the work of Langer \cite{langer}.
His approach has been applied to relativistic field theory by
Coleman \cite{coleman} and Callan \cite{colcal} and extended by
Affleck \cite{affleck} and Linde \cite{linde} to finite-temperature
quantum field theory. 
The basic quantity in this approach is 
the nucleation rate $I$, which
gives the probability per unit time and volume to nucleate a certain
region of the stable phase (the true vacuum) within the metastable 
phase (the false vacuum). 
The calculation of $I$ relies on a semiclassical approximation 
around a dominant saddle-point that is identified with 
the critical bubble. 
This is a static configuration 
(usually assumed to be spherically symmetric) within the metastable phase 
whose interior consists of the stable phase. 
It has a certain radius that can be determined from the 
parameters of the underlying theory. Bubbles slightly larger 
than the critical one expand rapidly, thus converting the 
metastable phase into the stable one. 

The nucleation rate is exponentially suppressed by the 
action of the critical bubble.
Possible deformations of the critical  bubble
generate a static pre-exponential factor.
The leading contribution to it
has the form of a ratio of fluctuation determinants and corresponds to the
first-order correction to the semiclassical result. 
For a four-dimensional theory of a real scalar field 
at temperature
$T$, the nucleation rate
is given by 
\be
I=\frac{E_0}{2\pi}
\left(\frac{\Gamma_b}{2\pi}\right)^{3/2}\left|
\frac{\det'[\delta^2 \Gamma/\delta\phi^2]_{\phi=\phibounce}}
{\det[\delta^2 \Gamma/\delta\phi^2]_{\phi=0}}\right|^{-1/2}
\exp\left(-\Gamma_b\right). 
\label{rate0} \ee
Here $\Gamma$ is the effective action 
of the system for a given configuration of the
field $\phi$ that acts as the order parameter of the problem. 
The action of the critical bubble is $\Gamma_b
=\Gamma\left[\phibounce(r)\right]-\Gamma[0]$,
where $\phibounce(r)$ is the spherically-symmetric
bubble configuration and $\phi = 0$ corresponds to the false vacuum. 
The fluctuation determinants are evaluated either 
at $\phi = 0$ or around $\phi=\phibounce(r)$. 
The prime in the fluctuation determinant around
the bubble denotes that the three zero eigenvalues 
of the operator $[\delta^2 \Gamma/\delta\phi^2]_{\phi=\phibounce}$
have been removed. 
Their contribution generates the factor 
$\left(\Gamma_b/2\pi \right)^{3/2}$ and the volume factor
that is absorbed in the definition of $I$ (nucleation rate per unit volume). 
The quantity $E_0$ is the square root of
the absolute value of the unique negative eigenvalue.

In field theory, the rescaled free energy 
density of a system for homogeneous configurations 
is identified with
the temperature-dependent effective potential. This is often evaluated 
through some perturbative scheme, such as the loop expansion~\cite{Col73}. 
In this way, the profile and the action of the bubble are determined
through the potential. This approach, however,
faces three fundamental difficulties:\\
a) The effective potential 
is a convex function of the field and seems inappropriate for the
 the study of tunnelling.\\
b) The fluctuation determinants in the expression for the nucleation
rate have a form completely analogous to the one-loop correction to
the potential. The question of double-counting the effect of
fluctuations (in the potential and the prefactor)
must be properly addressed. \\
c) The fluctuation determinants 
in the prefactor are ultraviolet divergent. 
An appropriate regularization scheme, consistent with the one used
in the calculation of the potential, must be
employed.

All the above issues can be resolved
through the implemention of the notion of coarse graining in the 
formalism \cite{coarse}, in agreement with Langer's philosophy.
The problem of computing the difference of the effective
action between the critical bubble and the false vacuum may be
divided into three steps:
In the first step, one only includes fluctuations with momenta
larger than a scale $k$ of the order of the typical gradients
of $\phi_b(r)$. For this step one can consider approximately
constant fields $\phi$ and use a derivative expansion for the resulting 
coarse-grained free energy 
$\Gamma_k[\phi]$. The second step searches for the
configuration $\phi_b(r)$ which is a saddle point of $\Gamma_k$.
The quantity $\Gamma_b$ in eq. (\ref{rate0}) 
is identified with $\Gamma_k[\phi_b]-\Gamma_k[0]$.
Finally, the remaining fluctuations with momenta smaller
than $k$ are evaluated in a saddle-point approximation
around $\phi_b(r)$. This yields the ratio of fluctuation
determinants with an ultraviolet cutoff $k$.
Langer's approach corresponds to a one-loop approximation
around the dominant saddle point
for fluctuations with momenta smaller than a coarse-graining
scale $k$. 

\subsection{Calculation of the nucleation rate}

In the following we review studies of nucleation 
based on the formalism of the effective average action
$\Gamma_k$, which can be identified with the 
free energy, rescaled by the temperature, at a given coarse-graining scale 
$k$. In the simplest case, we consider a
statistical system with one space-dependent degree of freedom described
by a real scalar field $\phi(x)$.
For example, $\phi(x)$ may correspond to
the density for the gas/liquid transition,
or to a difference in concentrations for chemical phase transitions,
or to magnetization for the ferromagnetic transition.
Our discussion also applies to a quantum field theory in
thermal quasi-equilibrium.
An effective three-dimensional description
applies for a thermal quantum field theory at 
scales $k$ below the temperature $T$. We assume that
$\Gamma_{k_0}$ has been computed (for example perturbatively) for $k_0=T$
and concentrate here on the three-dimensional
(effective) theory.

We compute $\Gamma_k$ by solving the flow equation
between $k_0$ and $k$. For this purpose we approximate $\Gamma_k$ by 
a standard kinetic term and
a general potential $U_k$. This corresponds to the 
first level of the derivative expansion \cite{indices,morris}.
The long-range 
collective fluctuations
are not yet important at a short-distance scale
$k_0^{-1}=T^{-1}$.
For this reason, we assume here a polynomial potential
\be
U_{k_0} (\phi) =
\frac{1}{2}m^2_{k_0} \phi^2
+\frac{1}{6} \gamma_{k_0} \phi^3
+\frac{1}{8} \lx_{k_0} \phi^4 \; .
\label{eq:two20} \ee
The parameters $m_{k_0}^2, \gamma_{k_0}$ and $\lambda_{k_0}$
depend on $T$ . 

We compute the form of the potential $U_k$ at scales $k\leq k_0$ by
integrating an evolution equation obtained from the exact
flow equation for the effective average action.
Here we employ a mass-like infrared cutoff $k$ for the fluctuations
that are incorporated in $\Gamma_k$. 
The evolution equation for the potential can be written as \cite{first,second}
\be
 \frac{\partial }{\partial k^2}\left[U_k(\phi)-U_k(0)\right]
= - \frac{1}{8 \pi}\left[\sqrt{k^2+U_k''(\phi)}
-\sqrt{k^2+U_k''(0)}\right]. 
\label{twofour} \ee
In this entire section primes denote derivatives with respect to $\phi$.

The form of $U_k$ changes as
the effect of fluctuations with momenta above the decreasing scale
$k$ is incorporated in
the effective couplings of the theory. 
In general, $U_k$ 
is not convex for non-zero $k$.
It approaches the convex effective potential only in the limit $k\to 0$.
In the region relevant for a first-order phase 
transition, $U_k$ has two distinct
local minima.
The nucleation rate should be computed for $k$ larger than
or around the scale $k_f$ at which  $U_k$ starts  receiving important
contributions from field configurations that interpolate between
the two minima. This happens when the negative curvature at the top
of the barrier becomes approximately equal to $-k^2$.
Another consistency check for the above choice of $k$  is  
the typical length scale of a thick-wall critical
bubble, which is $\gta  1/k$ for $k > k_f$.
The use of $U_k$ at a non-zero value of $k$ resolves the first fundamental
difficulty related to the convexity of the potential.

The other two difficulties are overcome as well.
In our approach the pre-exponential factor in eq. (\ref{rate0})
is well-defined and finite,
as an ultraviolet cutoff of order $k$ must implemented in the calculation of
the fluctuation determinants.
The cutoff must guarantee that fluctuations
with characteristic momenta $q^2 \gta k^2$ do not contribute to the
determinants. This is natural, as
all fluctuations with typical momenta above $k$ are
already incorporated in the form of $U_k$.

In order to implement the appropriate ultraviolet cutoff $\sim k$
in the fluctuation determinant, let us look at
the first step of an iterative solution of eq. (\ref{twofour})
\bea
U_k^{(1)}(\phi)-U_k^{(1)}(0)=&~&U_{k_0}(\phi)-U_{k_0}(0)
\nonumber \\
&+&
\frac{1}{2}\ln\left[\frac{\det[-\partial^2+k^2+U_{k}''(\phi)]}{
\det[-\partial^2+k^2_0+U_{k}''(\phi)]}
\frac{\det[-\partial^2+k^2_0+U_{k}''(0)]}{
\det[-\partial^2+k^2+U_{k}''(0)]}\right].
\label{iter} \eea
For $k\to 0$, this solution is a regularized one-loop
approximation to the effective
potential. Due to the ratio of determinants, only
momentum modes with $k^2<q^2<k_0^2$ are effectively included
in the momentum integrals. The form of the infrared cutoff in eq.
(\ref{twofour}) suggests that we should implement the ultraviolet cutoff
for the fluctuation determinant in the nucleation rate (\ref{rate0})
as
\bea
I
&\equiv&A_k \exp({-S_k})
\nonumber \\
A_k&=& \frac{E_0}{2\pi}\left(\frac{S_k}{2\pi}\right)^{3/2}
\left|
\frac{\det'\left[-\partial^2+U''_k(\phibounce(r)) \right]}
{\det \left[ -\partial^2+k^2  + U''_k(\phibounce(r)) \right]}
~\frac{\det\left[-\partial^2+k^2 + U''_k(0) \right]}
{\det\left[-\partial^2+U''_k(0)\right]}
\right|^{-1/2}.
\label{rrate} \eea
We switched the notation to $S_k=\Gamma_k[\phi_b]-\Gamma_k[0]$ instead
of $\Gamma_b$ in order to make the $k$-dependence in the
exponential suppression factor explicitly visible. 

As a test of the validity of the approach, the result for the rate
$I$ must be independent of the coarse-graining scale $k$, because
the latter should be considered only as a technical device.
In the following we show that this is indeed the case when
the expansion around the saddle point is convergent and the
calculation of the nucleation rate reliable. Moreover, the 
residual $k$ dependence of the rate can be used as
a measure of the contribution of the next order in the
saddle-point expansion. 

The critical bubble configuration $\phi_b(r)$ is an 
SO(3)-invariant solution of
the classical equations of motion that interpolates between 
the local maxima of the potential
$-U_k(\phi)$. It can be computed easily with numerical methods. 
The evaluation of the fluctuation determinants $A_k$ is more complicated. 
However, the regularized expression in (\ref{rrate}) can be computed through
a combination of numerical and analytical techniques \cite{first,second}.

\begin{figure}
\begin{center}
\hspace{0.cm}
\psfig{figure=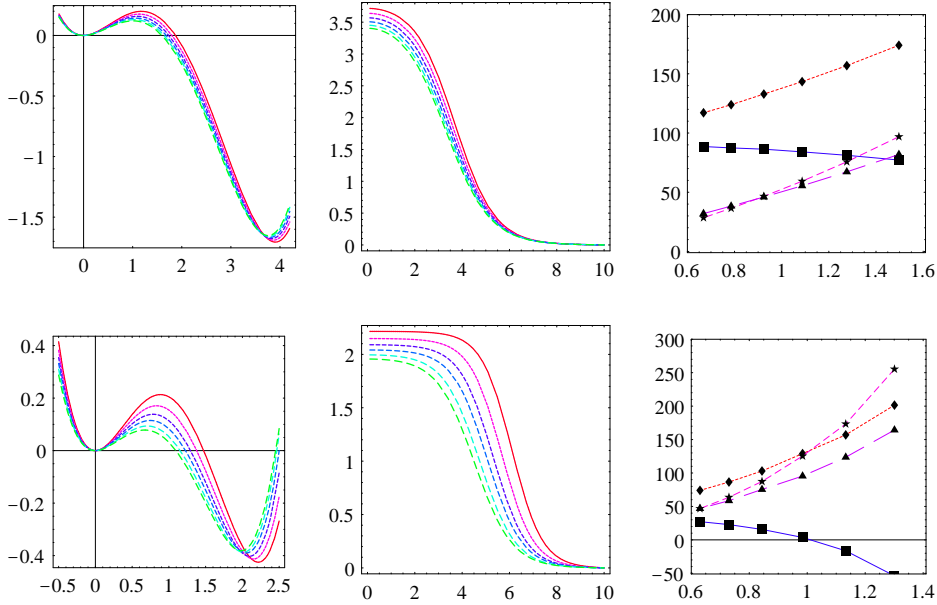,height=3.3in}
\end{center}
\caption{
Dependence of effective potential, critical bubble and
nucleation rate on the coarse graining scale $k$.
\label{fig:three}}
\end{figure}

\subsection{Two examples}

Sample computations are presented in fig. 3.
The first row corresponds to
a model with an initial potential given by eq. (\ref{eq:two20}) with
$m^2_{k_0}=-0.0433~k_0^2$, 
$\gamma_{k_0}=-0.0634~k_0^{3/2}$,
$\lx_{k_0}=0.1~k_0$. 
We first show the evolution of the potential $U_k(\phi)$ as the scale
$k$ is lowered. 
The solid line corresponds to $k/k_0=0.513$ while the
line with longest dashes (that has the smallest barrier height)
corresponds to $k_f/k_0=0.223$. At the scale $k_f$ the negative
curvature at the top of the barrier is slightly larger than
$-k_f^2$ and we stop the evolution.
The potential and the field have been
normalized with respect to $k_f$.
As $k$ is lowered from $k_0$ to $k_f$, the absolute minimum of the potential
settles at a non-zero value of $\phi$, while a significant barrier
separates it from the metastable minimum at $\phi=0$.
The profile of the critical bubble $\phi_b(r)$
is plotted in the second figure in units of $k_f$
for the same sequence of scales.  For $k\simeq k_f$ the characteristic
length scale of the bubble profile and $1/k$ are comparable. This is expected,
because the form of the profile is determined by the barrier of the potential,
whose curvature is $\simeq -k^2$ at this point.
This is an indication that we should not proceed to coarse-graining
scales below $k_f$.
We observe a significant
variation of the value of the field $\phi$ in the interior of the bubble
for different $k$.

Our results for the nucleation rate are presented in the third
figure.
The horizontal axis corresponds to $k/\sqrt{U''_k(\phi_t})$,
i.e. the ratio of the scale $k$
to the square root of the positive curvature (equal to the
mass of the field) at the
absolute minimum of the potential located at $\phi_t$.
Typically, when $k$ crosses below this mass, 
the massive fluctuations of the field
start decoupling. The evolution of the convex parts of
the  potential slows down and eventually stops.
The dark diamonds give the values of the action $S_k$ 
of the critical bubble. We observe a strong
$k$ dependence of this quantity.
The stars indicate the values of
$\ln ( A_k/k^4_f )$.
Again a substantial decrease with decreasing $k$ is observed. 
The dark squares give our results for
$-\ln(I/k^4_f )
= S_k-\ln ( A_k/k^4_f )$. It is remarkable that the
$k$ dependence of this quantity almost disappears for $k/\sqrt{U''_k(\phi_t})
\lta 1$.
The small residual dependence on $k$ can be used to estimate the
contribution of the next order in the expansion around the saddle point.
It is reassuring that this contribution is expected to be
smaller than $\ln ( A_k/k^4_f )$.

This behaviour confirms our expectation that the
nucleation rate should be independent of the scale $k$ that
we introduced as a calculational tool. 
It also demonstrates that
all the configurations plotted in the second figure give equivalent
descriptions of the system, at least for the lower values of $k$.
This indicates that the critical bubble should not be associated only
with the saddle point of the semiclassical approximation, whose
action is scale dependent. It is the combination of
the saddle point and its possible deformations
in the thermal bath that has physical meaning.

For smaller values of $|m^2_{k_0}|$ the dependence of the nucleation
rate on $k$ becomes more pronounced. We demonstrate this in the
second row in fig. 3 for which
$\lambda_{k_0}/(-m^2_{k_0})^{1/2}$ $=0.88$ 
(instead of $0.48$ for the first row).
The value of $\lambda_{k_0}$ is the same as before, whereas
$\gamma_{k_0}=-1.61\cdot 10^{-3}k_0^{3/2}$ and $k_f/k_0=0.0421$.
Higher-loop contributions to $A_k$ become important and the expansion
around the saddle point does not converge any more. There are two clear
indications of the breakdown of the expansion: \\
a) The values of the 
leading and subleading contributions to the nucleation rate, 
$S_k$ and $\ln ( A_k/k^4_f )$ respectively, become comparable. \\
b) The $k$
dependence of $\ln ( I/k^4_f )$ is strong and must be canceled
by the higher-order contributions.\\
The discontinuity in the order parameter 
at the phase transition is approximately 5 times
smaller in the second example
than in the first one. As a result, the second phase transition
can be characterized as weaker. Typically,
the breakdown of the saddle-point approximation
is associated with weak first-order phase transitions.

In the last figure of each row we also display the values of 
$\ln ( A_k/k^4_f )$ (dark triangles) predicted by the approximate expression 
\be
\ln \frac{A_k}{k^4_f}
\approx \frac{\pi k}{2} 
\left[
- \int_0^\infty \!\!\! r^3 \left[ 
U''_k\left( \phi_b(r) \right)
-U''_k\left( 0 \right)
\right] dr
\right]^{1/2}.
\label{eq:appr}
\ee
It gives a good approximation to the exact
numerical results, especially near $k_f$, and can be used for
quick checks of the validity of the expansion around the
saddle point.

\subsection{Region of validity of homogeneous nucleation theory}
\label{regionvalid}

It is useful to obtain some intuition on the behaviour of the 
nucleation rate by using the approximate expression~(\ref{eq:appr}).
We assume that the potential has a form similar
to eq.~(\ref{eq:two20}) even near $k_f$, i.e.
\be
U_{k_f} (\phi) \approx
\frac{1}{2}m^2_{k_f} \phi^2
+\frac{1}{6} \gamma_{k_f} \phi^3
+\frac{1}{8} \lx_{k_f} \phi^4.
\label{eq:two21} \ee
(Without loss of generality we take $m^2_{k_f}>0$.)
For systems
not very close to the endpoint of the first-order critical line,
our assumption is supported by
the numerical data.
The scale $k_f$ is determined by the relation
\be
k^2_f \approx \max \left| U''_{k_f}(\phi)
\right|=\frac{\gamma_{k_f}^2}{6\lx_{k_f}}-m_{k_f}^2.
\label{abc} \ee

Through the rescalings 
$r=\rt/m_{k_f}$, $\phi=2\phit\, m_{k_f}^2/\gamma_{k_f}$, the potential
can be written as 
$\Ut(\phit)=\phit^2/2-\phit^3/3+h\,\phit^4/18$, with
$h=9\lx_{k_f} m_{k_f}^2/\gamma_{k_f}^2$. 
For $h \approx 1$ 
the two minima of the potential have approximately
equal depth. The action of the saddle point can be expressed
as
\be
S_{k_f} = 
\frac{4}{9} \frac{m_{k_f}}{\lx_{k_f}}\, h \St(h),
\label{action2} \ee
where $\St(h)$ must be determined numerically through $\Ut(\phit)$.
Similarly, the pre-exponential factor can be estimated through
eq.~(\ref{eq:appr}) as
\bea
\ln \frac{ A_{k_f}}{k^4_f}  &\approx& 
\frac{\pi}{2}
\sqrt{ \frac{3}{2h}-1}\,\, \At(h),
\nonumber \\ 
\At^2(h) &=& 
\int_0^\infty \left[ \Ut'' \left(\phit_b(\rt)\right)-1 \right] \,\rt^3\, d\rt,
\label{prefactor2} \eea
with $\At(h)$ computed numerically.
Finally, the relative importance of the fluctuation
determinant is given by
\be
R=\frac{\ln \left(A_{k_f}/k_f^4 \right)}{S_{k_f}} \approx
\frac{9\pi}{8}\frac{1}{h}\sqrt{\frac{3}{2h}-1}\,
\frac{\At(h)}{\St(h)}
\,\frac{\lx_{k_f}}{m_{k_f}}
=T(h) \,\frac{\lx_{k_f}}{m_{k_f}}.
\label{fin} \ee
The ratio $R$ can be used as an indicator of the validity of 
the saddle point expansion. The latter is valid only for $R \lta 1$.

Numerical evaluation of 
$T(h)$ shows that it diverges for $h\to 0$, while it becomes 
small for
$h \to 1$.
This suggests two cases in which 
the expansion around the saddle point is expected to break down:\\
a) For 
fixed $\lx_{k_f}/m_{k_f}$, the ratio $R$ becomes larger than 1 for
$h\to 0$. In this limit the barrier becomes negligible and the 
system is close to the spinodal line. \\
b) For fixed $h$, $R$ can be large for sufficiently
large $\lx_{k_f}/m_{k_f}$. This is possible even for $h$ close
to 1, so that the system is far from the spinodal line. This 
case corresponds to weak first-order phase transitions, as 
can be verified by observing that 
the saddle-point action ~(\ref{action2}), 
the location of the true vacuum 
\be
\frac{\phi_t}{\sqrt{m_{k_f}}}=
\frac{2}{3}\sqrt{h}\,\,
\phit_t(h)\,\sqrt{\frac{m_{k_f}}{\lx_{k_f}}},
\label{phit} \ee
and the
difference in
free-energy density between the minima 
\be
\frac{\Delta U}{m_{k_f}^3}=
\frac{4}{9}\,{h}\,\,
\Delta\Ut(h)\,\frac{m_{k_f}}{\lx_{k_f}}
\label{deltaU} \ee
go to zero in the limit $m_{k_f}/\lx_{k_f}\to 0$ for fixed $h$. This
is in agreement with the discussion of fig. 3
in the previous subsection.\\
The breakdown of homogeneous nucleation theory in both the above
cases is confirmed through the numerical computation of the nucleation 
rates \cite{second}. 

\subsection{Other results}

The approach has been applied to more complicated systems,
such as theories of two scalar fields. 
The most interesting feature of the two-scalar models is 
the presence of radiatively induced first-order phase transitions.
Such transitions usually take place when 
the expectation value of a certain field generates the mass of
another through the Higgs mechanism.
The fluctuations
of the second field can induce the appearance of new minima in the 
potential of the first, resulting in first-order phase transitions.
We discussed such an example in the first section.
The problem of double-counting 
the effect of fluctuations is particularly acute 
in such situations. The introduction of a coarse-graining
scale $k$ resolves this problem, by separating the high-frequency 
fluctuations of the system which may be responsible for the 
presence of the second minimum through the Coleman-Weinberg mechanism
\cite{Col73},
from the low-frequency ones which are relevant for tunnelling. 

Unfortunately, the expansion around the saddle point does not converge
for radiatively induced first-order phase transitions. The 
prefactor resulting from the fluctuation determinant of the second field
is always comparable to the leading exponential term \cite{third}.
As a result, the saddle-point approximation breaks down
and the predicted nucleation rate $I/k^4_f$ is strongly $k$ dependent.
The above results are not surprising. The radiative corrections to the 
potential and the pre-exponential factor have a very similar form 
of fluctuation determinants. When the radiative corrections are large enough
to modify the bare potential and generate a new minimum, the 
pre-exponential factor should be expected to be important also. 
These findings have important
implications for cosmological phase transitions, such
as the electroweak \cite{third,fifth}.

Finally, the reliability of our approach has been confirmed 
through comparisons with results obtained through lattice simulations
\cite{fourth}, or alternative analytical methods in their region
of applicability \cite{sixth}.

\section*{Acknowledgments}
This research was supported in part by the European Commission under 
the RTN programs HPRN--CT--2000--00122, HPRN--CT--2000--00131 and
HPRN--CT--2000--00148.

\eject

\end{document}